\documentstyle[12pt,aaspp4,epsf]{article}

\begin{document}

\def\intldate{\number\day\space\ifcase\month\or
January\or February\or March\or April\or May\or June\or
July\or August\or September\or October\or November\or December\fi
\space\number\year}

%
%

\def \event	{{MACHO~97-BLG-28}}
\def \shortname	{{97-BLG-28}}
\def \deg    	{$^{\circ}$}
\def \etal   	{{et al.\thinspace}}
\def \eg     	{{e.g.,}}
\def \cf     	{{cf.}}
\def \ie     	{{i.e.,}}
\def \hub    	{$H_{\hbox{\rm 0}}$}
\def \hunits 	{km s$^{\hbox{\rm --1}}$ Mpc$^{\hbox{\rm --1}}$}
\def \kms    	{{\rm km~s$^{\hbox{\rm --1}}$}}
\def \sec    	{$^{s}$}
\def \arcsecpoint {{$^{''}\mkern-5mu.$}}
\def \asec	{{\arcsecpoint}}
\def \dophot	{D{\sc o}PHOT}
\def \DOPHOT	{D{\sc o}PHOT}
\def \hi   	{\ion{H}{1}}
\def\sb		{{\rm mag~arcsec$^{-2}$}}
\def\area	{${\rm deg}^2$}
\def\kpc	{\hbox{\rm kpc }}
\def\pc		{\hbox{ pc }}
\def\yr		{ \, {\rm yr}}
\def\peryr	{ \, {\rm yr^{-1} }}
\def\vlos	{ v_{\rm los} }
\def\lsim	{ \rlap{\lower .5ex \hbox{$\sim$} }{\raise .4ex \hbox{$<$} } }
\def\gsim	{ \rlap{\lower .5ex \hbox{$\sim$} }{\raise .4ex \hbox{$>$} } }
\def\solar	{ {\odot} }
\def\lsolar	{ {\rm L_{\odot}} }
\def\msolar	{ \rm {M_{\odot}} }
\def\rsolar	{ \rm {R_{\odot}} }
\def\surfmunit  { \rm {\, \msolar \, pc^{-2}} }
\def\HI		{{H{\sc I}}}
\def\mags	{{ \, \rm mag }}   
\def\percubicpc	{ { \pc^{-3} } }
\def\abs	{ \hbox{ \vrule height .8em depth .4em width .6pt } \,} 
\def \tightenlines {\def\baselinestretch{1}\small}
\def\gtorder	{\mathrel{\raise.3ex\hbox{$>$}\mkern-14mu\lower0.6ex\hbox{$\sim$}}}
\def\ltorder	{\mathrel{\raise.3ex\hbox{$<$}\mkern-14mu\lower0.6ex\hbox{$\sim$}}}

%
%
%

\vskip 2cm

\title{
Limb-Darkening of a K Giant in the Galactic Bulge:\\
\vskip 0.15cm
PLANET Photometry of MACHO 97-BLG-28}

\vskip 1cm
\author{M.D. Albrow\altaffilmark{1,2}, 
J.-P. Beaulieu\altaffilmark{3},
J. A. R. Caldwell\altaffilmark{1},\\ 
M. Dominik\altaffilmark{3,4,5}, 
J. Greenhill\altaffilmark{6}, 
K. Hill\altaffilmark{6},
S. Kane\altaffilmark{4,6}, 
R. Martin\altaffilmark{7},\\  
J. Menzies\altaffilmark{1}, 
R. M. Naber\altaffilmark{3}, 
J.-W. Pel\altaffilmark{3},  
K. Pollard\altaffilmark{1,2},\\ 
P. D. Sackett\altaffilmark{3}, 
K. C. Sahu\altaffilmark{4}, 
P. Vermaak\altaffilmark{1},  
R. Watson\altaffilmark{6}, 
A. Williams\altaffilmark{7}}
\author{(The PLANET Collaboration)}
\author{and M. S. Sahu\altaffilmark{8,9}}
\affil{}

\altaffiltext{1}{South African Astronomical Observatory, P.O. Box 9, 
Observatory 7935, South Africa}
\altaffiltext{2}{Univ. of Canterbury, Dept. of Physics \& Astronomy, 
Private Bag 4800, Christchurch, New Zealand}
\altaffiltext{3}{Kapteyn Astronomical Institute, Postbus 800, 
9700 AV Groningen, The Netherlands}
\altaffiltext{4}{Space Telescope Science Institute, 3700 San Martin Drive, 
Baltimore, MD~21218~~U.S.A.}
\altaffiltext{5}{Universit\"at Dortmund, Institut f\"ur Physik, 44221 Dortmund,
Germany}
\altaffiltext{6}{Univ. of Tasmania, Physics Dept., G.P.O. 252C, 
Hobart, Tasmania~~7001, Australia}
\altaffiltext{7}{Perth Observatory, Walnut Road, Bickley, Perth~~6076, Australia}
\altaffiltext{8}{NASA/Goddard Space Flight Center, Code 681, Greenbelt, MD~ 20771~~U.S.A.}
\altaffiltext{9}{National Optical Astronomy Observatories, 950 N. Cherry Avenue,
Tucson, AZ~87519-4933~~U.S.A.}
\vskip 1cm


\newpage 

\begin{abstract}

We present the PLANET photometric dataset$^*$ for the 
binary-lens microlensing event \event\ consisting of 696 $I$ and $V$-band 
measurements, and analyze it to determine the radial 
surface brightness profile of the Galactic bulge source star. 
The microlensed source, demonstrated to be a K giant by 
our independent spectroscopy, crossed an isolated cusp of 
the central caustic of the lensing binary, generating a sharp peak in the 
light curve that was well-resolved by dense (3 --- 30 minute) and 
continuous monitoring from PLANET sites in Chile, South Africa, and Australia. 
This is the first time that such a cusp crossing has been observed. 
Analysis of the PLANET dataset has produced a    
measurement of the square root limb-darkening coefficients 
of the source star in the $I$ and $V$ bands; the resulting 
stellar profiles are in excellent agreement 
with those predicted by stellar atmospheric models for K giants.  
The limb-darkening coefficients presented here are the first derived 
from microlensing.  They are also among the first for normal giants 
by any technique and the first for any star as distant as the 
Galactic bulge.  
Modeling of our light curve for \event\ indicates that the 
lensing binary has a mass ratio $q = 0.23$ and an (instantaneous) 
separation in units of the angular Einstein ring radius 
of $d = 0.69$.  For a lens in the Galactic bulge, this corresponds 
to a typical stellar binary with a projected separation between 
1 and 2~AU.  If the lens lies closer (\ie\ in the Galactic disk), the 
separation is smaller, and one or both of the lens objects is in 
the brown dwarf regime.  Assuming that the source is a bulge K2 giant 
at 8~kpc, the relative lens-source proper motion is 
$\mu = 19.4\, \pm 2.6 \, \mbox{km}\,\mbox{s}^{-1}\,\mbox{kpc}^{-1}$, 
consistent with a disk or bulge lens.  If the non-lensed blended 
light is due to a single star, it is likely to be a young white 
dwarf in the bulge, consistent with the blended light coming 
from the lens itself. 

\end{abstract}


\vskip 2cm

\hangindent 2cm{\small 
$^*$Based on observations at:  Canopus Observatory, Tasmania, Australia; 
the European Southern Observatory, La~Silla, Chile;  
and the South African Astronomical Observatory, Sutherland, South Africa.}


\newpage

\section{Introduction} \label{intro}

The PLANET (Probing Lensing Anomalies NETwork) collaboration 
uses a longitudinally-distributed network of telescopes  
in the southern hemisphere to perform densely-sampled, precise 
photometric monitoring of Galactic microlensing events 
with the express goal of searching for anomalies, 
especially those that may betray the presence of extrasolar 
planets (\cite{planetpilotpaper}) orbiting the lenses.      
Multiple lenses, such as lensing binary stars or planetary systems, 
generate measurably anomalous light curves if the background 
source comes close to a caustic generated by the lens;  
direct transits of a caustic can cause sudden and dramatic  
changes (amplification $\gtorder$ 10 on time scales of hours) 
in the apparent brightness of the source.  
Three such microlensing 
events were monitored by PLANET in its 1997 Galactic bulge 
observing season; here we describe our results for one of these,
\event.

Although typical subgiants and giants in the Galactic bulge  
have radii that subtend only  $\sim$1$ - $10~microarcseconds ($\mu$as), 
the dramatic change in magnification in the vicinity of caustics can 
selectively magnify the spatial structure of the source star by huge factors, 
momentarily acting as a very high-resolution, large-aperture telescope 
trained on the background source.  These relatively rare but spectacular  
lensing events are particularly important because they can yield otherwise 
unobtainable information about the lens and source.  

Generally speaking, the only physical parameter 
that can be deduced from non-anomalous microlensing light curves 
is the characteristic time scale $t_{\rm E}$ of the event, 
defined as the time required for the lens to move an 
angular distance across the observer's sight line to the source equal 
to the angular Einstein radius $\theta_{\rm E}$  
\begin{equation}
\theta_{\rm E} = \sqrt{\frac{4GM}{c^2}\,\frac{D_{\rm LS}}{D_{\rm L}\,D_{\rm{S}}}}\,,
\end{equation}
where $M$ is the total mass of the lens and $D_{\rm L}$, $D_{\rm S}$, 
and $D_{\rm LS}$ denote the observer-lens, observer-source and 
lens-source linear distances, respectively.  

In caustic crossing events, photometric monitoring can allow 
more information to be obtained, including the relative proper motion $\mu$ 
of the lens-source system and --- if the coverage is 
complete over the caustic --- the surface brightness profile of the 
source star.  
During a caustic transit, the radial surface brightness profile, or limb-darkening, of the source influences the shape of the 
resulting light curve (\cite{witt95}).  
Photometric data of sufficient precision and temporal resolution 
can thus be inverted to deduce the limb-darkening profile 
of source stars as distant as the Galactic bulge (\cite{bc96}; \cite{sassiap}),  including possible chromatic effects 
induced by differential limb-darkening at different 
wavelengths 
(\cite{bc95a}, 1995b; 
\cite{vallsgabaudchromo95}, 1998; 
\cite{gouldwelch96}).  
Time-resolved spectroscopic monitoring with very large aperture telescopes 
may reveal even more information, including detailed information 
about stellar atmospheric physics 
(\cite{loebsasse95}; \cite{gaudi99}; \cite{heyrovsky99}) 
and metallicity (\cite{lennon96}) of the sources. 

One of the most important observational constraints on  
models of stellar atmospheres has come from limb-darkening measurements 
of the Sun, a relatively cool dwarf.   
Constraints on the atmospheres of other types of stars are more 
difficult to obtain.  Measurements of radial profiles, or more specifically, 
limb-darkening parameters, have been attempted for only a handful of stars 
(see \cite{scholz97} and references therein).  
Eclipsing binaries, lunar occultations and interferometry 
have been used to resolve and measure the surface structure of nearby stars. 
 
The classical method of using the light curves of eclipsing binaries 
has provided accurate limb-darkening measurements or 
consistency checks in only a few cases (Popper 1984; Andersen 1991; 
Ribas \etal\ 1987), but at least one 
suggests that a non-linear limb-darkening law is indicated 
(Milone, Stagg \& Kurucz 1992).   
Direct HST imaging of the bright, nearby M2 supergiant Betelgeuse has 
resulted in estimates for its limb-darkening and other surface irregularities  
(\cite{gildupresstars96}). 
Interferometric techniques also require very large, bright stars 
(Hanbury Brown \etal\ 1974; Dyck \etal\ 1996; Mozurkewich \etal\ 1991; 
Mourard \etal\ 1997), 
where surface structure is often due to spotting or dust envelopes rather than 
limb-darkening (Roddier \& Roddier 1985; Wilson \etal\ 1992).  
Although the interferometric signal is 
often incompatible with a uniform stellar disk, 
with the exception of extreme supergiants like Betelgeuse  
(\cite{cheng86}; \cite{burns97}), 
theoretical limb-darkening laws are adopted in order to arrive at 
better stellar radius determinations 
(Quirrenbach \etal\ 1996; Hajian \etal\ 1998). 
Nevertheless, the technique holds promise for limb-darkening 
determinations as it begins to probe into the normal K giant regime 
(\cite{dyck98}).   Lunar occultations have been used to 
resolve stellar surface structure (Bogdanov \& Cherepashchuk 
1984, 1990, 1991; Richichi \& Lisi 1990; Di~Giacomo \etal\ 1991), 
but the dust associated with these supergiants complicates 
limb-darkening estimates here as well 
(Richichi \etal\ 1991, 1995) and surface profiles from theory are 
generally assumed so as to derive better diameters 
(White \& Feierman 1987; Richichi \etal\ 1998).

The precision of microlensing in mapping stellar surface brightness 
profiles can thus make an important contribution 
to stellar atmospheric theory by providing limb-darkening and spectral 
information on stars too faint or too small to be studied by 
traditional methods.  
We report here our precise and temporally dense photometric 
observations of the MACHO-detected microlensing event 97-BLG-28, 
which when combined with our spectral typing and modeling 
have produced the first limb-darkening measurement 
via microlensing, and the first resolution of   
surface structure in a star so distant ($\sim$8~kpc).  

\section{PLANET Observations of \event}\label{data}

The Galactic bulge microlensing event \event\ was alerted by the 
MACHO Collaboration on 29 May 1997\footnote{MACHO alerts are posted  
at http://darkstar.astro.washington.edu} 
as a normal event expected to reach peak brightness near 10 June 1997.   
PLANET began to observe the event directly after the initial alert 
as part of its normal monitoring program.  
Within hours of the anomalously rapid increase in its brightness 
on 14 June 1997, PLANET\footnote{PLANET anomaly alerts are posted  
at http://www.astro.rug.nl/$\sim$planet}  
and MACHO/GMAN independently issued electronic 
alerts announcing \shortname\ as a likely caustic-crossing binary event. 
Dense photometric PLANET monitoring continued over the peak 
of the light curve and for about 6 weeks thereafter, with additional 
data taken near the baseline magnitude in October 1997 and at the 
beginning of the 1998 season.  
A pre-caustic spectrum was taken on 31 May 1997. 

\subsection{Photometric Monitoring}

Three PLANET telescopes participated in the monitoring: 
the Dutch/ESO 0.92~m at La~Silla, Chile, 
the SAAO~1m at Sutherland, South Africa, 
and the Canopus~1m near Hobart, Tasmania. 
The full PLANET dataset consists of 686 data points taken in 1997 
(508 I-Band: 247 La~Silla, 128 SAAO, 133 Tasmania; 
178 V-Band: 98 La~Silla, 38 SAAO, 42 Tasmania) and another 
10 data points (5 I-Band; 5 V-band) collected at the 
SAAO~1m in March and April of 1998.  
For details concerning the telescopes, detectors, and field sizes, 
refer to Albrow \etal\ (1998).  
The data were reduced 
with \dophot\ (\cite{dophot}) 
using fixed position catalogs and four stars in the 
field chosen to be stable, moderately bright and relatively uncrowded 
as a relative flux standard (\cite{planetpilotpaper}).   
Instrumental magnitudes were calibrated against contemporaneous 
observations of Johnson-Cousins, UBV(RI)c E-region 
standards (Menzies \etal\ 1989) made at SAAO at the beginning of the 
1998 season.  
As we describe in \S3.1, individual 
zero points for photometry at different PLANET sites allow us 
to refer all our multi-site data to the SAAO natural system, and thus 
to the standard system.  Measurements taken in 1998 and 
calibrated in the manner described above produced 
baseline magnitudes of the source star of Johnson 
$V = 17.91$ and Cousins $I = 15.66$, accurate to 0.05~mag. 

Whenever the event was 
fainter than $I \approx 14.7$, measurements derived from frames 
with poor image quality (FWHM $> 2.2\arcsec$) were clearly and systematically 
fainter than those from moderate to high quality images, 
and had a scatter that was substantially larger than the 
\dophot-reported error bars.  
In order to avoid systematically biasing our modeling, 
measurements with $I \gtorder 14.7$ (corresponding to 
$V \gtorder 17$) and FWHM$ > 2.2\arcsec$ were excluded from the fitting 
procedure to extract lens and source parameters.  
In addition, one particularly poor image that produced 
measurements that deviated by 0.75~mag for constant stars 
as bright as $I = 14$ was also excluded.  This high-quality subset of our 
data\footnote{This PLANET dataset for \event\ is publicly available   
at http://www.astro.rug.nl/$\sim$planet} consists of 586 data points 
(431 I-Band: 247 La~Silla, 130 SAAO, 54 Tasmania; 
155 V-Band: 98 La~Silla, 41 SAAO, 16 Tasmania) and is displayed in Fig.~1.
 
\subsection{Spectroscopy}

On 31 May 1997, prior to the anomalous peak of \event,  
red and blue spectra were taken of the source star by Dr. M.~Sahu 
using the high-throughput EFOSC (ESO Faint Object 
Spectrographic Camera) on the ESO~3.6m at La Silla, Chile.  
The camera has both imaging and spectroscopic capabilities, and can hold 
5 gratings simultaneously so that multiple spectral resolutions and 
bandpasses can be easily configured.

A direct image of the field was first obtained with an exposure of 10~sec. 
The 512 $\times$ 512 pixel Tektronix CCD provided a 
5.2$\arcmin \times$ 5.2$\arcmin$ field of view.  
No filter was used in order to minimize any possible offset between the
slit and the direct image.  
A slit of 1\arcsecpoint5 width was used for all the observations, 
and was placed in the correct parallactic angle. 
Observations were conducted through thin cirrus clouds 
in seeing of $\sim$1\arcsecpoint0.   
The blue spectrum was obtained at 04:40 UT using the B150 grating, 
which covers 3900-5300 \AA, with a resolution of about 5\AA.   
The red spectrum was obtained at 05:25 UT
using the O150 grating, which covers the wavelength range 5200 - 6800 \AA,
with the same resolution.  
The exposure time in both cases was 2400 sec.

The image processing software packages MIDAS and IRAF were used to
reduce the spectral data.  After bias
subtraction, the frames were flat-field corrected with an average
of 5 flat-field images of the dome illuminated with a tungsten lamp.
To ensure good sky subtraction, sky levels were determined from 
both sides of the spectrum.  The resulting one-dimensional spectrum 
was then wavelength calibrated with a He-Ar spectral lamp 
and flux calibrated against the standard star LTT~4364.  
Shown in Fig.~2 is the resulting geocentric (lab frame), 
combined spectrum for the source star of \shortname.  
Obvious cosmic rays have been removed. 
The spectral resolution is $\sim 5$\AA, 
corresponding to a velocity resolution of 
about 260 \kms\ at the strong NaI absorption feature near 5890\AA. 
The heliocentric correction for the velocities is 9.5 \kms, which 
is negligible compared to the resolution of the spectrum.

\subsection{Radius and Spectral Type of the Source}

Comparison of the flux-calibrated spectrum shown in Fig.~2 with 
spectral synthesis codes ({\sc FITGRID} and {\sc SPECFIT} tasks of IRAF) 
yields the best match to a K2III star 
of solar metallicity, effective temperature $T=4350 \,$K and 
surface gravity ${\rm log} ~ g=1.9$.  The strong NaI line 
is not typical of a star of this type, but the equivalent 
width of the line is entirely consistent with interstellar absorption along 
this long sight line to the bulge source star at celestial (J2000) 
coordinates 
$\alpha = \, $18$^{\rm h}$00$^{\rm m}$33.8$^{\rm s}$ and 
$\delta = \, -$28\deg01$\arcmin$10$\arcsec$, 
corresponding to Galactic $l = 2.46$\deg\ and $b = - 2.36$\deg. 
Some ambiguity remains in the spectral solution depending on the 
extinction to the source, but the best fit for the reddening of the 
spectrum and the radius of the source are 
consistent with that derived from an independent analysis 
of the color-magnitude diagram of field, described below. 

The radius of the source star was determined by considering its position in 
the $(V-I)_{0} - I_{0}$ color-magnitude diagram (CMD) relative to 
red clump (RC) stars in the Galactic bulge.  Our CMD was derived 
from observations taken at the SAAO~1m, calibrated by reference to 
E-region standard star baseline data in 1998.  

The absolute $I$ magnitude of the peak of the RC luminosity function in 
the bulge has been determined to be $I_{RC} = -0.23 \pm$~0.03 
(Stanek \& Garnavich 1998; Paczy\'nski \& Stanek 1998). 
We have fitted the $I$-magnitude distribution of the RC stars in our sample 
using the same function as Stanek \& Garnavich (1998). 
Based on the distribution of stars in CMD, 
we find that there is an equal probability that the 
\shortname\ source star is in 
the RC or on the red giant branch (RGB). 
Possible values for the average field reddening were
determined by fitting the observed RGB to the 
Bertelli \etal\ (1994) isochrones for ages of 3.1 and 10 Gyr and
metallicities [Fe/H] = $-$0.4, 0, +0.4, resulting in the range of 
reddening values $0.8 < E(V-I)  < 1.2$.  
Stanek's (1998) reddening map of the Galactic plane 
indicates a gradient of $\sim$ 0.2 mag across the $3 \times 3$~arcmin
SAAO field;  interpolation at the position of the microlensing event 
gives $E(B-V) = 0.88$, corresponding to $E(V-I) = 1.10$.

For each value of plausible reddening, we used the known RC 
absolute magnitude to rescale our CMD, 
treating the distance modulus as a free parameter. 
For each isochrone fit to the RGB for a given age and metallicity, other 
isochrones were also plotted to allow for the possibility that the 
source star may differ in age or metallicity
from the RGB mean.  In all cases for which a reasonable fit  
($\Delta(V-I)_{0} < 0.1$~mag) was found 
for the source being on the RGB or in the RC, 
the source radius was calculated from the appropriate isochrone.  
Reasonable consistency emerged among the 25 resulting radius
values, with all in the range $13 < R_*/R_\odot < 20$, 
and a mean value of $R_*/R_\odot = 15 \pm 2$ (1-$\sigma$ uncertainty).  
This result will 
be used together with the modeling in \S\ref{params} to estimate the 
relative proper motion of the lens with respect to the source.

In Fig.~3, we show our CMD for the field of \shortname, dereddened 
with our fitted value of $E(V-I) = 1.095$ ($A_I = 1.56$) 
over the whole of the field.  
The position of the observed (blended) source star at baseline is shown as the 
filled circle in Fig.~3, where it is assumed that the 
source suffers the same reddening as the field mean.   
Our derived dereddened color of $(V-I)_0 = 1.16$ 
from CMD considerations is in excellent agreement with the 
$(V-I)_0 = 1.14$ (Bessell 1990) expected for our K2 giant spectral typing 
of this source.

\section{An Extended-source, Binary-lens Model for \event}\label{model}

The abrupt change of slope in the light curve near 
HJD - 2449719 = 896.4 evident in Fig.~4 suggests that the source star 
has crossed the caustic structure of a binary lens.  
The single nearly-symmetric finite peak near HJD - 2449719 = 895.6 
further suggests that the extent of the source is resolved 
by the observations and that the source passed behind an isolated cusp 
of the caustic geometry on a nearly perpendicular trajectory.
This lensing geometry is indeed borne out by our modeling of the light curve 
for \shortname, as can be seen in Fig.~5.  To reproduce the features 
in the observed light curve it was necessary to include 
lens binarity, source blending, extended source size, and 
source limb-darkening in this model.  Together with 
allowances for possible photometric offsets between the 
3 observing sites in both bands, this requires a total of 19 fit parameters, 
as we now explain.

We label each subset of the data by observing site and filter.  
Let $k$ denote the number of such data subsets.  
A model with a point-source and point-lens will require 
$3+k$ parameters:
$t_{\rm E} = \theta_{\rm E}/\mu$, the (Einstein) time required for 
the lens to move an angular distance $\theta_{\rm E}$ (Eq.~1); 
$u_0$, the smallest lens-source angular separation in units 
of $\theta_{\rm E}$; 
$t_0$, the time at which this smallest separation occurs; and the set of $k$  
$m_0^i$, the multi-site, multi-band baseline magnitudes.
Individual baselines are required for each site because although 
all data of a given band are referred to the same secondary field 
standards, we often find that small discrepancies remain in our 
multi-site photometry that appear to be related to differences 
in detector resolution and average seeing conditions (Albrow \etal 1998).

Parametrization of a binary lens composed of two objects with mass 
$M_1$ and $M_2$ requires an additional 3 parameters:
$d$, the angular binary separation in units of $\theta_{\rm E}$; 
$q = M_2/M_1$, the binary mass ratio; and 
$\alpha$, the angle between the line from $M_2$ to $M_1$ and
the direction of source proper motion 
(relative to the lens).  
These parameters and the direction of proper motion are chosen so 
that $u_0 \geq 0$, $0 \leq \alpha < 2\pi$, $0 < q \leq 1$, 
and the midpoint of the lens system is on the right hand side 
of the moving source as viewed by the observer.  
For a binary system, the smallest angular separation $u_0$ 
between lens and source now refers to the midpoint of the lens system, and
the angular Einstein radius $\theta_{\rm E}$ refers to the total mass 
$M = M_1 + M_2$.

The baseline magnitudes include all light, including any that may 
be due to any unresolved, non-lensed light that may be confused with 
source light in the observations.  To account for the possibility of 
such photometric confusion, blending parameters $f_j$ ($j =1\ldots{}n$) 
must be added for each of the observed $n$ wave bands
to characterize the fraction of the total light contributed 
by the source alone at baseline. 
No blending in wave bands $j$ corresponds to $f_j = 1$.
The extended size of the source is characterized by an additional 
parameter $\rho_\ast \equiv \theta_\ast/\theta_{\rm E}$, 
the angular radius of the source in units of the Einstein radius.
Finally, the light intensity profile of the source may be limb-darkened 
rather than uniform over the stellar disk.
We adopt a limb-darkening law of the form
\begin{equation}
I_\lambda(\vartheta) = I_\lambda (0) \left[1-c_\lambda \, (1-\cos \vartheta) - d_\lambda \, (1-\sqrt{\cos \vartheta}
)\right]\,, 
\end{equation}
where $\vartheta$ 
is the angle between the normal to the stellar surface and the 
line of sight.  
The coefficients $c_\lambda$ and $d_\lambda$ are wavelength dependent, 
requiring an additional $2n$ parameters.

The inclusion of all these effects thus requires $7+k+3n$ fit parameters. 
For the 3 PLANET sites, each observing \event\ in $V$ and $I$,  
$k = 6$ and $n = 2$, thus necessitating 19 parameters in the full model.

\subsection{Fitting the Model to the Data}

We fit three different extended-source binary-lens models, with either a
uniform source or a 1- or 2-parameter limb-darkened source, 
to the combined, high-quality dataset of 586 CCD frames (\S2.1) 
using the formal uncertainties reported by DoPHOT. Details of calculating
the light curves for such models can be found in the work of Dominik (1998). 
The multi-site, multi-band baselines were allowed to 
vary independently in the fitting process;  
this photometric alignment resulted in relative multi-site offsets of 
$0.007 - 0.065 \, $mag in our best model.     

In general, it is a difficult task to find a minimum in high-dimensional
parameter space.  However, in this case, nearly optimal values for 
some parameters could be found by first searching in lower-dimensional
subspaces.  The six baselines could be estimated from the latest data points
and then fitted with a point-lens model together with the parameters
$u_0$, $t_0$, $t_{\rm E}$ and the two blending parameters, using the
data points outside the peak region.  We thus began 
fitting the binary-lens extended-source models 
only after we had good guesses for 11 of the parameters.  
Parameters from the uniform source fit then provided us with 
good initial estimates for a total of 15 parameters, requiring only 
2 or 4 additional parameters when including limb-darkening.  

Our best fit was achieved for a 2-parameter limb-darkening model (LD2),
and yielded a $\chi^2_{\rm min}$ of 1913 for the 
567 degrees of freedom (d.o.f).  
This model is displayed in the left panels of Fig.~4 and its  
fit parameters are listed in Table~\ref{fpartab}.   
If the LD2 model is indeed the best 
representation of the data, then the reduced 
$\chi^2_{\rm min} / {\rm d.o.f.}\, = 3.374$ 
is an indication that the \dophot\ uncertainties are on average 
underestimated by 
a factor $\sim$1.8, consistent with our previous experience 
(\cite{planetpilotpaper}).     

The corresponding source trajectory and caustic structure for the 
LD2 model shown in
Fig.~5 illustrate that the cusp caustic of the binary lens just 
sweeps over the limb of the extended source, the first clear observation of 
such a cusp crossing.  Such a crossing differentially magnifies 
different portions of the source as a function of time (top panel, Fig.~6). 
Because the central ring crosses 
almost directly over the cusp, it experiences the greatest magnification.  
The fraction of light contributed by concentric rings (of equal area) 
on the stellar disk therefore also varies during the anomaly 
(bottom panel, Fig.~6), allowing 
the surface profile, and thus limb-darkening, of the source to be 
deduced.  For the specific geometry of \shortname, the contrast 
between the fraction of light contributed by the inner and 
outermost of 10 equal-area rings is a factor $\sim$3 during the cusp crossing. 
The PLANET dataset is well-sampled everywhere during the cusp 
crossing except in the 12-hour period near ${\rm HJD - 2449719 = 895}$.

\section{Limb-Darkening of the Bulge K Giant Source Star}\label{limbdark}

In addition to our best fitting model with square root limb-darkening 
coefficients (LD2), 
Table~\ref{fpartab} also lists the parameters for two other models: 
a fit with a linear, 1-parameter, limb-darkening law (LD1),
which yields $\chi^2_{\rm min}$ of 1930 for 569 d.o.f., and a 
fit with a uniformly bright source (UB), which yields 
$\chi^2_{\rm min}$ of 3255 for 571 d.o.f.   
Under the assumption that our 2-parameter limb-darkening model is 
indeed the best representation of the data, the large $\chi^2$ compared to the number of degrees of freedom must be attributed to a misestimation 
of the experimental uncertainties.  This is not unexpected since 
our previous experience in crowded fields indicates that \dophot\ 
formal uncertainties are smaller than the true scatter in the photometry 
of constant stars by an amount that depends on the crowding of 
the star (Albrow \etal\ 1998).  In order to assess conservatively 
the differences in $\chi^2_{\rm min}$ between models, therefore, we first 
rescale the formal \dophot\ uncertainties by 
$\sqrt{1913/567} = \sqrt{3.374} \approx 1.8$, 
which is tantamount to the assumption that the LD2 model is perfect 
and would yield a true $\chi^2_{\rm min} = 567$ if the ``true'' photometric 
uncertainties were known.  The rescaling is global, preserving the 
relative uncertainty between points, and conservative, allowing for 
the possibility that part of the $\Delta\chi^2_{\rm min}$ between models 
can be due to the underestimation of photometric uncertainties by \dophot.  
The rescaled $\chi^2_{\rm min}$ 
of the LD1 and UB models are thus 572 and 965, respectively.  

The significance of the limb-darkened LD1 and LD2 models over the 
uniform brightness UB model is enormous: using the rescaled 
uncertainties, $\Delta \chi^2_{\rm min} = 393$ between the LD1 and UB models, and $\Delta \chi^2_{\rm min} = 398$ between LD2 and UB.  
This precipitous drop in $\chi^2_{\rm min}$ with 
the addition of the limb-darkening parameters leaves 
no doubt that limb-darkening has been detected to an exceedingly high 
degree of confidence in the PLANET photometry of this event.
The enlargements of the light curve in Fig.~4 show how precisely the  
2-parameter limb-darkening model (LD2) reproduces the detail in the 
cusp and limb regions in both the $I$ and $V$ bands.  
On the right of the same figure, these sections of the light curve are 
overplotted with the uniform source model (UB).   
The multi-site photometry is aligned automatically 
by the fitting procedure; this photometric alignment 
is somewhat different for the two models.   
Even given the freedom to photometrically realign the data from 
different sites relative to one another, the UB model fails to fit the 
shape of the convexity at peak and the gentle slope of the curve 
as the limb of the star egresses.  The failure of the uniformly 
bright model to reproduce the structure in the light curve during 
the limb crossing can be seen most dramatically in Fig.~7, where 
the residuals of the LD2 and UB models are shown.

If the smaller normalized $\chi^2_{\rm min}$ of the LD2 model 
compared to the LD1 models is not physically significant, but instead 
only due to the freedom of adjusting the additional $d_\lambda$ parameters, 
we would expect 
$\Delta \chi^2_{\rm min}$ to be distributed as $\chi^2$  
with a number of degrees of freedom equal to the number of additional 
parameters (Press \etal\ 1986, Chapter 14).   
The significance of the LD2 model over the LD1 
model is thus determined by the 8.2\% probability of obtaining 
by chance a $\Delta \chi^2_{\rm min} > 5$ improvement  
with two additional degrees of freedom, corresponding to a   
marginal 1.7$\sigma$ detection of a surface profile that deviates 
from the 1-parameter limb-darkening law in both bands.  
This marginal improvement indicates that the inclusion of additional  
profile fitting parameters beyond those included in the LD2 model 
is unlikely to result in a significantly better fit; 
the LD2 model contains the all the information 
that we are able to pull from this dataset about the source profile.  

The large difference in $\Delta \chi^2_{\rm min}$ between the 
limb-darkened and uniform bright models is not due solely to the 
cusp-crossing portions of the light curve.  
Since the model parameters are correlated, even those parameters 
unrelated to the source profile differ between the uniform source model 
and the limb-darkened models, as inspection of Table~\ref{fpartab} 
indicates.  
Interestingly, the limb-darkened models always required a smaller 
photometric offset between sites in both bands.  The uniform 
brightness model apparently attempted to match the observed shape 
of the light curve at the limb by adjusting the photometric offsets slightly, 
hindering the $\chi^2$ performance of the UB model elsewhere 
in the light curve.  
The clear signature of limb-darkening is revealed only  
with high precision data during the few hours when the limb of the 
star is grazing the caustic; high-quality data over the whole of the 
light curve is required, however, to obtain a full and accurate 
microlensing solution.

\subsection{Robustness of the Measurement}

Another test of the reliability of our limb-darkening parameters 
is provided by fits we performed on the full dataset, including the poorer 
quality frames, but now estimating the photometric uncertainties empirically 
so that the size of the error bar scales with the overall frame quality.  
Specifically, we set the uncertainty of the event magnitude on a given frame 
equal to the average deviation of all similarly-bright constant stars 
from their average magnitude.  
We also eliminated the baseline points taken in 1998 in order to test 
whether the availability of a long temporal baseline was important to 
our conclusions about limb-darkening.  The resulting $\chi^2_{\min}$ and 
fit parameters are given in the last three columns of Table~\ref{fpartab} 
for the three source models.  The value of $\chi^2_{\min}$ 
now appears to be too small primarily due to the fact that 
\shortname\ is less crowded than a typical star of its brightness, 
so that the frame quality undercertainties are overestimates of its 
true photometric scatter. 
What is important to note is that the uniformly bright source is 
again strongly ruled out and that all model parameters 
are left almost entirely unaffected by this alternate selection and 
treatment of the data, demonstrating the model's robustness.

Both of the simpler models are special cases of the LD2 family of models: 
the LD1 model requires the limb-darkening parameter $d_\lambda = 0$ in both  
$I$ and $V$, while the UB model sets 
$c_\lambda = d_\lambda = 0$ in both bands.  This means that our LD1 and UB 
models, which correspond to points on the $\chi^2$ hypersurface 
over restricted portions of the LD2 parameter space, also correspond 
to points on the full LD2 $\chi^2$ hypersurface.  
The probability of obtaining a renormalized 
$\chi^2$ that is larger than that of 
the LD1 solution is $> 10\%$ for both the 
\dophot\ and frame quality methods of estimating relative photometric 
uncertainties; the probability of obtaining a 
renormalized $\chi^2$ larger than that of the UB is negligibly small.

We conclude that the 2-parameter limb-darkening model is clearly superior 
to the uniform source model and marginally superior to the 1-parameter 
limb-darkening model.  We now adopt it as our best model and use it 
to derive the physical parameters for the lens and the  
surface brightness profile of the source, which we now  
compare to expectations from stellar atmosphere models.  

\subsection{Comparison to Stellar Atmospheric Models} 

As we discussed in \S\ref{data}, both spectroscopic and photometric 
considerations indicate that the source star of \shortname\ is 
a K giant in the Galactic bulge, with K2III being the most probable typing. 
The fits to our photometric data discussed in \S\ref{model}
yield the limb-darkening coefficients $c_\lambda$ and $d_\lambda$ for 
the $V$ and $I$ bands separately.   
The corresponding surface brightness profiles for the source star, 
normalized so as to give a total flux equal to unity, are shown 
in Fig.~8 for our LD2 
model of the high-quality photometric dataset using 
\dophot\ estimates for the relative photometric uncertainties.  
The surface brightness profiles derived from modeling the full dataset 
with empirical ``frame quality'' error bars produces nearly identical profiles 
(see coefficients in Table~1).  Our results clearly indicate that, as expected, 
this giant is more limb-darkened in $V$ than in $I$.  

Also shown in Fig.~8 are predictions from stellar atmospheric models 
for K0 and K5 giants from two different groups 
(\cite{vanhammelimbdark}; \cite{diazlimbdark}; \cite{claretlimbdark}) 
both using the square-root limb-darkening law of Eq.~2.  
We have interpolated between the values given by 
these authors, labeling as K0 a star with effective temperature 
$T = 4750 \,$K and surface gravity ${\rm log} \, g = 2.15$, and as K5 
a giant $T = 4000 \,$K and surface gravity ${\rm log} \, g = 1.75$, in 
good agreement with standard convention (\cite{lang}).    
The cooler, later type K5 
giants are more limb-darkened than their warmer K0 cousins. 
Although the theoretical $I$ profiles of both groups are nearly identical, 
the $V$ profiles of Diaz-Cordoves, Claret \& Gimenez (1995) 
are slightly more limb-darkened than those of van~Hamme (1993).  

The surface brightness profiles derived from our photometric data 
modeled with the limb-darkening law of Eq.~2 
are in excellent agreement with these predictions from atmospheric 
models.  Although Fig.~8 indicates that the source star of \shortname\ 
may be slightly more limb-darkened in the $I$ band than current models 
predict for its spectral type, this conclusion should be treated with 
considerable caution given the uncertainties in the light curve modeling, 
spectral typing, and the rather ad hoc nature of the fitting 
function for the surface brightness profile.  

Our results for \shortname\ represent the first time that 
limb-darkening coefficients have been measured for a microlensing 
source.  
Limb-darkening was indicated but not measured by MACHO observations of  
the high-amplitude single-lens event MACHO~95-BLG-30; fixing 
limb-darkening coefficients to agree with stellar atmosphere models did  
produce a fit to the data that was marginally better 
(normalized $\Delta \chi^2 = 3.4$) than uniform disk models 
(\cite{alcockmb9530}).  
Our limb-darkening measurement for \shortname\ is among only a very few 
determinations for normal giants, and respresents the first 
surface brightness profile for any star as distant as the Galactic bulge.   
The ability of PLANET 
photometry to yield measurements of limb-darkening coefficients 
in \event\ is due to the overall characterization of the complete 
light curve combined with excellent temporal coverage 
(sampling times $\ltorder 30$ minutes) over peak and limb caustic egress 
regions.

\section{Physical Parameters of the Lens}\label{params}

In caustic crossing events, dense photometric monitoring allows 
not only the determination of $t_{\rm E}$, the time required for the 
lens to travel an angular distance $\theta_{\rm E}$, but also 
a second time scale: the time $t_\ast$ required 
to travel the angular source radius $\theta_\ast$. 
For uniform rectilinear motion, the ratio $t_\ast/t_{\rm E}$ directly yields  
the angular size of the source in units of the Einstein ring, 
$\rho_\ast \equiv \theta_\ast/\theta_{\rm E}$.  
If $\theta_\ast$ can be estimated by other means   
(\eg\ photometric or spectroscopic typing), both 
$\theta_{\rm E}$ and the relative lens-source proper motion  
$\mu \equiv \theta_{\rm E}/t_{\rm E}$ can thus be determined, 
yielding important information on lens kinematics 
(\cite{gouldfinsource94}; \cite{nemwick94}; 
\cite{wittmao94}; \cite{peng97}). 
To date, proper motions derived from this technique 
have been published for only two binary events in the Magellanic Clouds, 
MACHO~LMC-09 (\cite{alcock97lmcevents}; \cite{bennettonlmc9}) 
and MACHO~98-SMC-01 (\cite{erossmc9801paper}; \cite{planetsmc9801paper};  \cite{machosmc9801paper}; \cite{oglesmc9801paper}; 
\cite{mpssmc9801paper}), and one 
single-lens (point-caustic) event MACHO~95-BLG-30 in the Galactic bulge 
(\cite{alcockmb9530}).  

The relative proper motion can be written in terms of the model 
parameters $t_{\rm E}$ and $\rho_{\ast}$, and the physical source size $R_\ast$ 
and distance $D_S$ as: 
\begin{equation}
\mu = \frac{R_{\ast}}{D_{\rm S}\,\rho_{\ast}\,t_{\rm E}}\,,
\end{equation}
which for $D_{\rm S} = 8.0~\mbox{kpc}$ and $R_{\ast} = 15~R_{\sun}$, as 
indicated by the spectral typing of \S2.3,  
yield a relative proper motion 
\begin{equation}
\mu = 19.4\,
\left(\frac{R_{\ast}}{15~R_{\sun}}\right)\,\mbox{km}\,\mbox{s}^{-1}\,
\mbox{kpc}^{-1} = 
11.2\,\left(\frac{R_{\ast}}{15~R_{\sun}}\right)\,\mu\mbox{as}\,
\mbox{day}^{-1}\,.
\end{equation}
This corresponds to a perpendicular velocity in the lens plane
\begin{equation}
v = x \, D_{\rm S}\,\mu = 155 \, x \, 
\left(\frac{R_{\ast}}{15~R_{\sun}}\right)\,\mbox{km}\,\mbox{s}^{-1}\,, 
\end{equation}
where $ x \equiv D_{\rm L}/D_{\rm S}$.  
The (total) mass of the lens is given by 
\begin{equation}
M = \frac{c^2}{4G D_{\rm S}}\,\frac{R_{\ast}^2}{\rho_{\ast}^2}\,
\frac{x}{1-x} = 0.09\,\left(\frac{R_{\ast}}{15~R_{\sun}}\right)^2\,M_{\sun} \frac{x}{1-x}\,,
\end{equation}
so that for a disk lens halfway to the Galactic Center ($x=0.5$), 
one obtains $v = 78~\mbox{km}\,\mbox{s}^{-1}$ and total lens mass 
$M = 0.09~M_{\sun}$, whereas a lens embedded in the Bulge with $x=0.9$ 
yields $v = 140~\mbox{km}\,\mbox{s}^{-1}$ and
$M = 0.81~M_{\sun}$.  Either value is consistent with Galactic kinematics. 

Assuming a source distance $D_S = 8 \,$kpc and source radius 
in the range $13 - 17~R_{\odot}$ (\S2.3), the top panel of Fig.~9 gives the 
mass of the individual lensing components as a function of $x$.  
Since the binary mass ratio is $q = 0.234$, it is quite likely that 
the lighter of the lens objects has the mass of an M dwarf or less; 
if the lens lies in the Galactic disk with $x < 0.8$, at least the smaller 
of the binary components is a brown dwarf.   
If, on the other hand, the lensing binary resides in the bulge 
$x > 0.8$, the masses 
of its components would be consistent with that of a typical 
lower main-sequence 
binary with a projected separation $a_p = R_{\ast} x d / \rho_{\ast}$ 
between 1 and 2~AU (bottom panel, Fig.~9).  
Disk lenses would have smaller separations.

\subsection{Blended Light: Light from the Lens Itself?}

Our model fits to \shortname\ yield a fraction $1 - f$ of blended 
(non-lensed) light that is larger in the $V$ than in the $I$ band, 
indicating that the blend star is very much bluer 
than the background source star (Table~1).  If the blend is due to a 
single star, it lies in an underpopulated portion of the CMD (Fig.~3), 
either because it is not at the mean field distance and has been 
improperly dereddened, or because it is in a short-lived phase of its 
evolution.   A main sequence star with $(V-I)_0 \ltorder 0.6$ 
has $I_0 < 3.7$, and thus is so intrinsically bright that it cannot appear 
as the blend anywhere along the reddening vector of Fig.~3 
unless it lies substantially 
{\it behind\/} the bulge, where it is likely to have more reddening 
than the \shortname\ field, not less.  

If, on the other hand, the blend star is somewhat  
in front of the bulge and experiences slightly less extinction, 
it could lie on the portion of the isochrone at $(V-I)_0 = - 0.4$ 
that corresponds to the transition 
zone between planetary nebulae and white dwarfs.   
Such an interpretation would be consistent, furthermore, with the blended 
light coming from the lens itself: a white dwarf mass of 
$M_{\rm WD} \approx 0.55 \, M_\odot$ is consistent with the 
mass that the larger lens component ($M_1$) would have at a distance  
$x \approx 0.88$, slightly in front of the bulge.  This dense portion of 
the Milky Way is a likely region for microlenses and should 
have an extinction similar to (but slightly less than) 
that of the field mean (see Fig.~3).  If this explanation is correct, 
then the smaller lensing component would have a mass just above 
the hydrogen-burning limit at $M_2 \approx 0.13  \, M_\odot$. 
These conclusions are valid only if the blended light 
sensed by the light curve fitting is due to a single star, 
but any explanation requiring two stars to produce the color 
of the blend would be even more ad hoc.   


\section{Conclusions} \label{conclude}

The temporal coverage of the bulge microlensing event \event\ from 
three PLANET observing sites longitudinally distributed in the southern hemisphere has resulted in an excellent characterization of the 
light curve in the $I$ and $V$ bands.  
The full data set consists of 513 photometric measurements in the $I$-band 
and 183 measurements in $V$, with 586 of these being of very high quality.   
During a 30-hour period over the peak of the light curve, 
sampling times between $\sim 3$ and 30 minutes were continuously 
maintained from the three sites.  Less dense photometric monitoring 
over other portions of the light curve, combined with baseline 
measurements taken nearly a year later have resulted in the 
best-characterized microlensing light curve to date.  
Our spectrum of \shortname, taken at moderate magnification, indicates  
that the source star is likely to be a highly reddened K2 giant of 
solar metallicity.  This conclusion is supported by a comparison of 
the color-magnitude diagram of the field to theoretical isochrones.

Modeling of the PLANET photometric dataset for \shortname\ 
clearly indicates that the disk of the 
resolved source is transited by the cusp of the central caustic 
generated by a binary lens with mass ratio $q = 0.23$ and 
instantaneous separation, in units of the Einstein ring radius,  
of $d = 0.69$.  
A source of uniform surface brightness is strongly ruled out by 
our data;  the derived square root limb-darkening coefficients in the 
$I$ and $V$ bands derived from the photometric data alone are in 
excellent agreement with expectations from stellar atmospheric models 
for K giants.  Under the assumption that the source star is a K2III, 
the models appear slightly more limb-darkened in $I$ than our 
measurements, but the difference may not be significant.  

All of the conclusions above are independent of the (unknown) 
lens-source distance ratio $x$ and (rather well-constrained) source 
distance $D_S$ and radius $R_{\ast}$.   Assuming a source in the  
Galactic bulge at $8 \,$kpc of radius $R_{\ast} = 15 \pm 2 \, R_\odot$, 
as indicated by our spectral typing, the total mass of the binary  
and its projected separation can be determined.  
If the lens resides in the bulge, it is likely to be a lower main sequence  
binary with both components above the hydrogen burning limit 
and a projected separation between 1 and 2~AU.  
If the lens resides in the disk with the 
lens-source distance ratio $x < 0.8$, the projected separation would  
be smaller and one or both of the components would have a mass 
in the brown dwarf regime.  
Assuming the same source distance and radius, the relative 
lens-source proper motion derived from our modeling is  
$\mu = 19.4\, \pm 2.6 \, \mbox{km}\,\mbox{s}^{-1}\,\mbox{kpc}^{-1}$ or
$\mu = 11.2 \pm 1.5 \, \mu$as/day, consistent with the lens having 
disk or bulge kinematics.  The uncertainty in the proper motion is 
dominated by the uncertainty in the source radius, not 
by the characterization of the microlensing light curve.  
If the unlensed, blended light indicated by our models is due to a 
single star, its color and magnitude suggest that it is a
young white dwarf in the bulge.  This conclusion is consistent with 
the light coming from the lens itself: the larger lens component 
would have a typical white dwarf mass at a bulge distance 
of $x \approx 0.88$, in which case its companion would 
be a very faint late M dwarf.

These results mark the first time that the surface structure of 
a source star has been measured via microlensing and the first 
limb-darkening determination of a star as distant as the Galactic Bulge 
by any technique.  As the number of caustic-crossing events alerted 
in real time continues to rise, and the quality and frequency of 
photometric monitoring of microlensing light curves continues to 
improve, microlensing holds the promise of making a substantial 
contribution to the field of stellar atmospheres by determining the 
surface profiles of normal stars too faint or small to be easily 
measured by other techniques.


\acknowledgments

PLANET thanks the MACHO collaboration for providing the 
original real-time electronic alert of this event; such alerts are 
crucial to the success of our intensive microlensing monitoring.  
We are grateful to observers from the Astronomical Society of Tasmania, 
especially Bob Coghlan, who made many of the key observations 
reported here.  We thank Andy Gould for reading a penultimate 
version of this manuscript. 
Financial support from Nederlands Wetenschapelijk Onderzoek, through award 
ASTRON 781.76.018, is gratefully acknowledged.  
PLANET members also wish to thank The Leids Sterrewacht Foundation, 
the South African Astronomical Observatory, Canopus Observatory, 
the European Southern Observatory, and Perth Observatory for the 
generous allocations of time that make our work possible.
The work of M.D. has been financed by a research grant from the Deutsche Forschungsgemeinschaft while at STScI, Baltimore and by a 
Marie Curie Fellowship at Kapteyn Institute, Groningen.



\clearpage 

\begin{deluxetable}{lccccccc}
\tablewidth{0pt}
\tablecaption{PLANET model parameters for Galactic bulge event \event. 
\label{fpartab}}
\tablehead{\colhead{ } & \multicolumn{3}{c}{Formal \dophot\ Error Bars}
 & \colhead{ } & \multicolumn{3}{c}{``Frame Quality'' Error Bars} \nl
\colhead{} & 
\colhead{LD2} & \colhead{LD1} & \colhead{UB} & \colhead{ } & 
\colhead{LD2} & \colhead{LD1} & \colhead{UB}}
\startdata
$\chi^2_{\rm min}$	& 1913   & 1930   & 3255  &~& 264    & 281    & 450 \nl
\# d.o.f		&  567   &  569   &  571  & & 667    & 669    & 671 \nl
\# parameters		&   19   &   17   &   15  & &  19    &  17    &  15 \nl
$t_{\rm E}$ [d]		& 27.3   & 27.2   & 26.1  & & 27.3   & 27.3   & 26.1\nl
$t_0$ [d]		& 895.58 & 895.58 & 895.63 && 895.58 & 895.58 &895.63\nl
$u_0$			& 0.0029 & 0.0030 & 0.021 & & 0.0029 & 0.0034 & 0.021\nl
$d$			& 0.686  & 0.687  & 0.678 & & 0.686  & 0.688  & 0.680\nl
$q$			& 0.234  & 0.232  & 0.277 & & 0.234  & 0.231  & 0.276\nl
$\alpha$		& 1.426  & 1.424  & 1.406 & & 1.427  & 1.422  & 1.419\nl
$f_{\rm I}$		& 0.97   & 0.98   & 0.98  & & 0.97   & 0.98   & 0.96 \nl
$f_{\rm V}$		& 0.84   & 0.84   & 0.85  & & 0.85   & 0.88   & 0.87 \nl
$\rho_\ast$		& 0.0286 & 0.0287 & 0.0296 && 0.0286 & 0.0289 &0.0295\nl
$c_{\rm I}$		& 0.40   & 0.75   & 0     & & 0.38   & 0.83   & 0 \nl
$c_{\rm V}$		& 0.55   & 0.92   & 0     & & 0.55   & 0.95   & 0 \nl
$d_{\rm I}$		& 0.37   & 0      & 0     & & 0.38   & 0      & 0 \nl
$d_{\rm V}$		& 0.44   & 0      & 0     & & 0.43   & 0      & 0 \nl
$m_0 ^{\rm LI}$    & 15.632 & 15.633 & 15.620 & & 15.630 & 15.631 & 15.597 \nl
$m_0 ^{\rm SI}$    & 15.643 & 15.644 & 15.631 & & 15.643 & 15.645 & 15.609 \nl
$m_0 ^{\rm TI}$    & 15.708 & 15.703 & 15.727 & & 15.697 & 15.697 & 15.673 \nl
$m_0 ^{\rm LV}$    & 17.899 & 17.898 & 17.889 & & 17.909 & 17.930 & 17.900 \nl
$m_0 ^{\rm SV}$    & 17.906 & 17.905 & 17.903 & & 17.914 & 17.938 & 17.903 \nl
$m_0 ^{\rm TV}$    & 17.926 & 17.918 & 17.965 & & 17.947 & 17.959 & 17.968 \nl
\enddata
\tablecomments{
In separate columns, model parameters for 2-parameter limb-darkened (LD2),
1-parameter limb-darkened (LD1) and 
uniformly bright (UB) source models are listed. 
The first three columns indicate fits of high quality data with 
\dophot-reported error bars; the last three fits are of 
all data with empirical ``Frame Quality'' uncertainties based on 
typical scatter for similar brightness stars on individual frames.  
The baselines $m_0$ refer to the total modeled flux (source + blend) 
at the given observing site (L=LaSilla, S=SAAO, T=Tasmania) and in 
the given band (I or V). 
}
\end{deluxetable}
 

\clearpage

\section*{Figure Captions}

\figcaption{
{\it Left:\/} The high-quality subset of PLANET photometry for 
\shortname\ spanning a 300 day period and 
consisting of 431 $I$-band points (bottom) and 155 
$V$-band points (top) from three PLANET observing stations in 
Tasmania (green), Chile (black), and South Africa (red) is shown 
with baseline points taken the following season.  
Light curves for our best-fitting limb-darkened model (LD2) are 
superimposed.  
{\it Right:\/} An enlargement showing the 30-day period near the peak.}

\figcaption{Flux-calibrated, combined red and blue spectra taken with 
EFOSC on the ESO 3.6m.  The spectrum has not been dereddened, and was taken 
before the anomalous peak when the event was only slightly magnified.}

\figcaption{
Calibrated, dereddened color-magnitude diagram of the \event\ field, 
taken from SAAO~1m PLANET data.  A reddening of  $E(V-I) = 1.095$ 
has been applied, 
as spectral typing and isochrone fitting to the red clump and red giant branch.   Also shown are the dereddened position on the diagram 
of the unmagnified, blended source (filled circle), and 
the positions of the blend and unblended source (open circles) 
as inferred from modeling. 
Isochrones from Bertelli \etal\ 1994 of age 6~Gyr and solar and super-solar 
metallicity are overplotted.   
The faint blend star would lie along indicated reddening vector 
(i.e., would be fainter and redder than shown) if it suffered less 
extinction than the field mean.
}

\figcaption{{\it Left:\/}
Light curves for our best fit, the 2-parameter limb-darkening model (LD2),  
superimposed on PLANET $V$ and $I$ data sets for a 3-day period centered 
on the cusp-crossing (bottom) and a 16-hour period during which 
the stellar limb swept over the cusp (top).  
{\it Right:\/} Same for the uniform bright source model (UB). 
\dophot\ estimates for relative photometric uncertainties have been assumed;   
the rescaled error bars are generally comparable to the size of 
the plotted points, and so have been omitted.}

\figcaption{
Configuration of binary lens and trajectory of the extended source on the sky 
from the 2-parameter limb-darkening model.  
The lens caustics are shown as the closed, pointed shapes in the 
center and right side of the diagram. 
The path of the source is indicated by the straight parallel lines; 
the distance between the two outer parallel lines corresponds to 
the source width.  
Lens positions are shown as crosses; $M_1$ is the heavier of the two.  
All distances are in units of the angular Einstein radius. 
}

\figcaption{{\it Top:\/} 
Magnification curves for 10 uniformly bright, equal-area rings
of a source crossing a cusp with the geometry of Fig.~5.  
{\it Bottom:\/} The contribution to the total light at any given 
time from each of the uniformly bright rings.
}

\figcaption{$V$ and $I$ band residuals (Model - Data)
for the models of Fig.~4 during the brief period in    
which the cusp swept over the trailing limb of the source.  
(Rescaled) formal \dophot\ error bars are shown.  The different  
sampling rates and observing conditions at the three  
sites (successively La~Silla, Tasmania and SAAO) is apparent.  
{\it Left:\/} Residuals for our best fitting 2-parameter 
limb-darkening model (LD2).  
{\it Right:\/} Same for the uniformly bright source model (UB). 
}

\figcaption{
Source radial profiles in the I-band ({\it left\/}) and V-band ({\it right}) 
for uniformly-bright (thin solid line) 
and our 2-parameter limb-darkened source (thick solid line) models.  
Also shown are the theoretical expectations 
in the Cousins I and Johnson V bands based on 
atmospheric models for K0 --- K5 giants (dashed lines) from two 
different literature sources (see text).  
}

\figcaption{
{\it Top panel:\/} 
The larger ($M_1$, solid lines) and smaller ($M_2$, dashed lines) 
lens masses shown as a function of $x$, the ratio of 
lens to source distances.  
Parallel lines enclose the range of uncertainty dominated by the 
uncertainty in the source radius, estimated to lie between 
13 and 17$\,R_{\sun}$.
A source distance of $8 \,$kpc has been assumed.  
{\it Bottom panel:\/} 
The projected binary separation is shown in AU as a function of $x$ for 
the same source radii.}

\newpage

\vglue -2cm
\hglue -2cm
\epsfxsize=18cm\epsffile{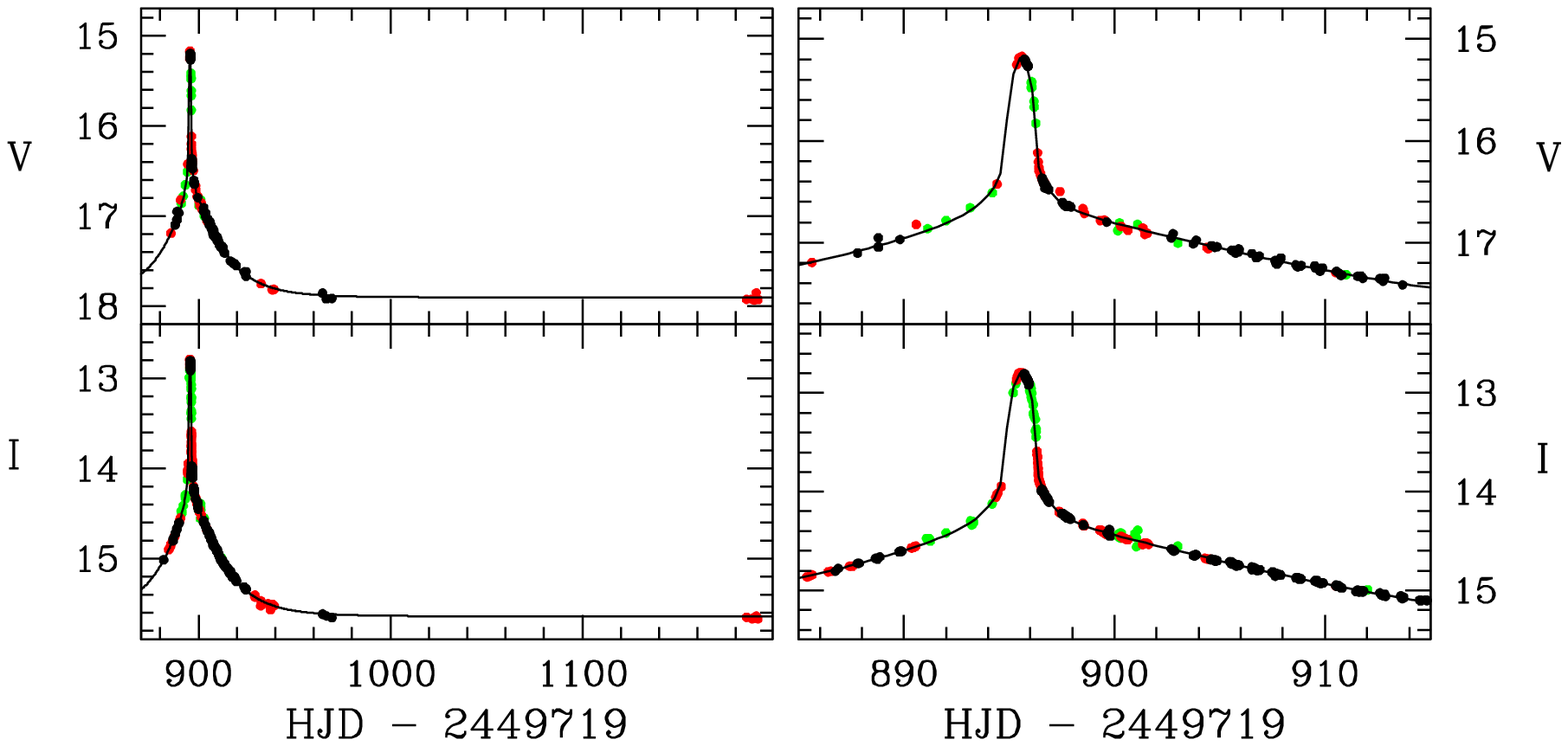}

\newpage

\vglue 1cm
\epsfxsize=15cm\epsffile{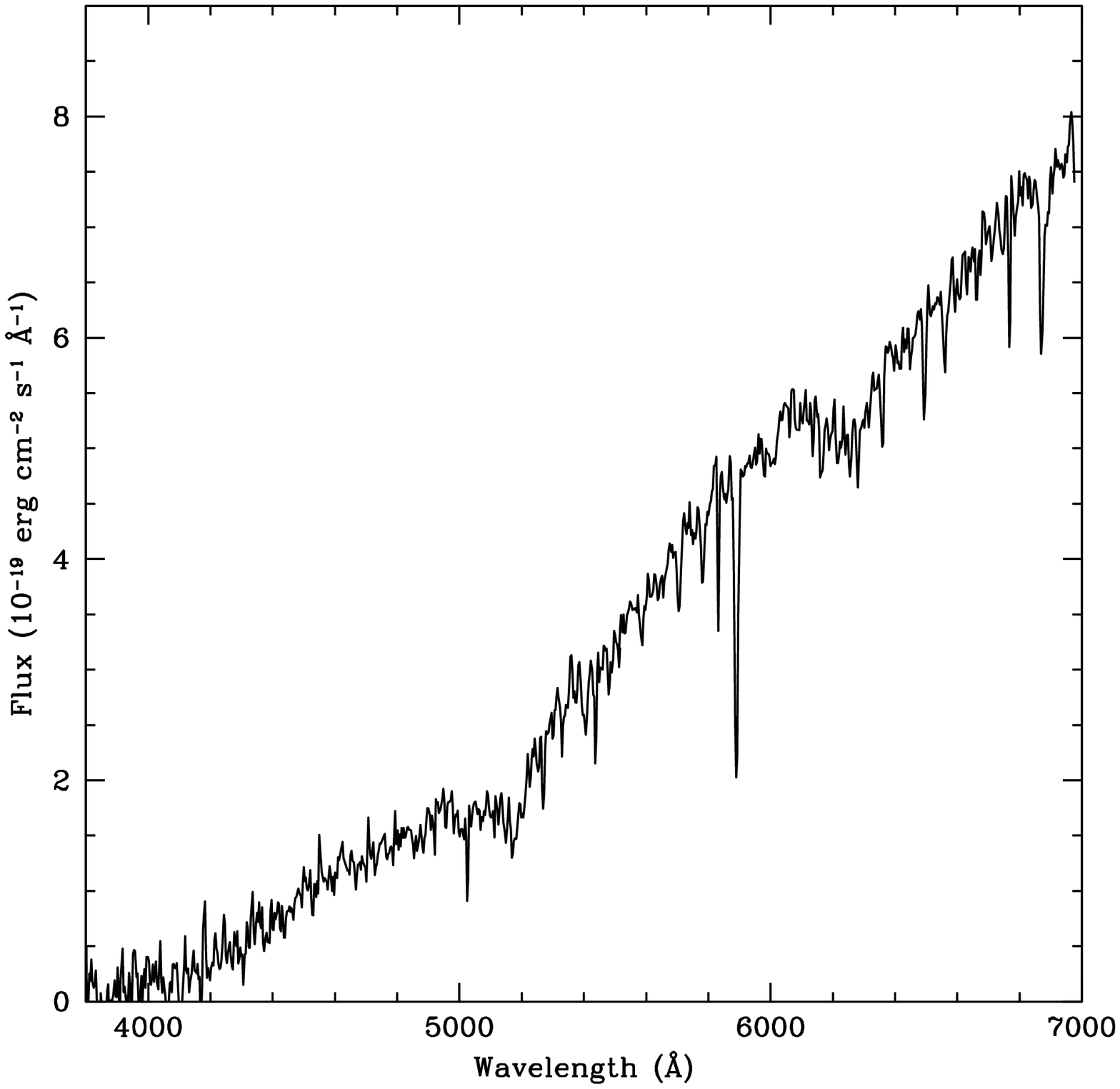}

\newpage

\vglue 1cm
\hglue -2cm
\epsfxsize=15cm\epsffile{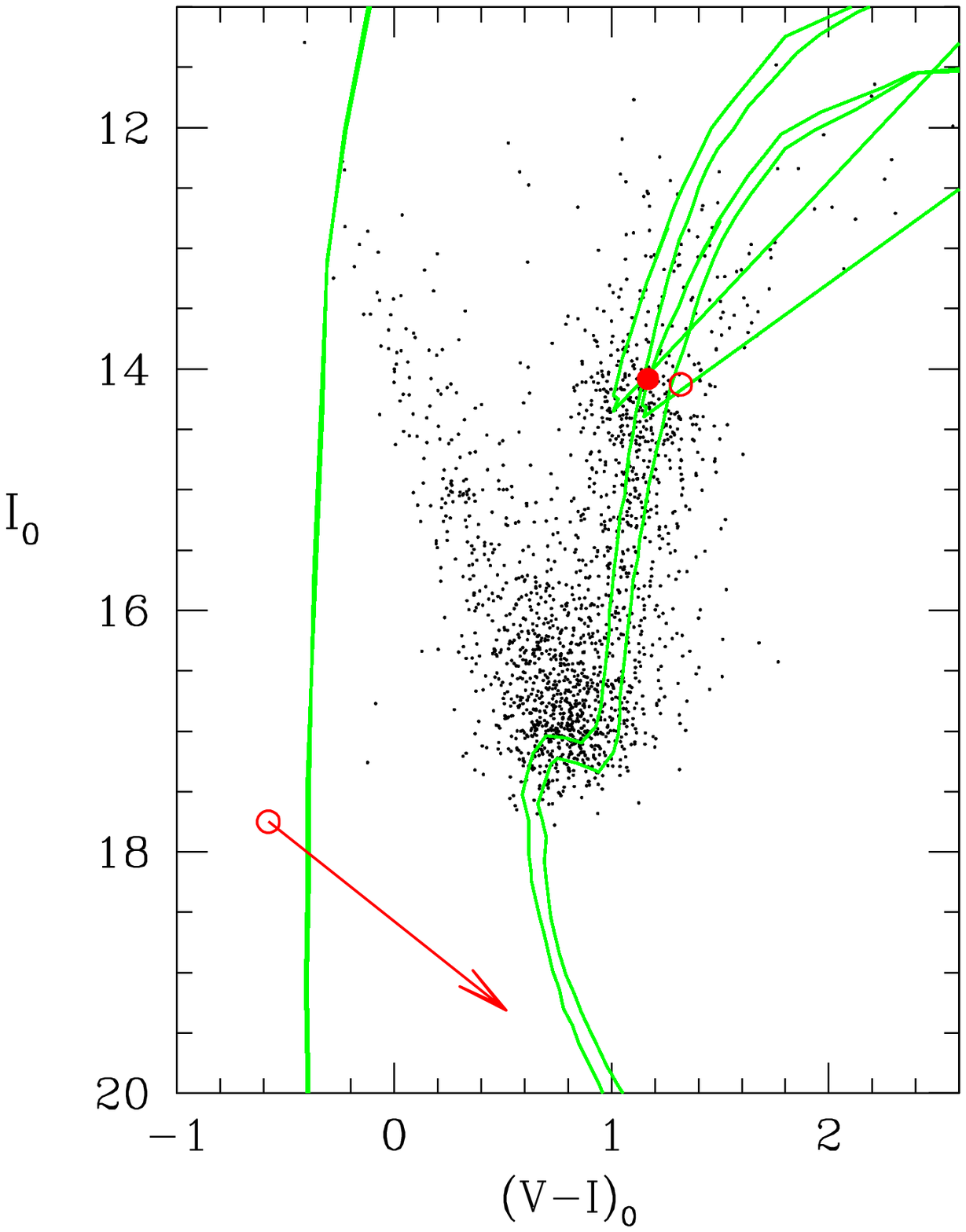}

\newpage

\vglue 1cm
\hglue -2cm
\epsfxsize=18cm\epsffile{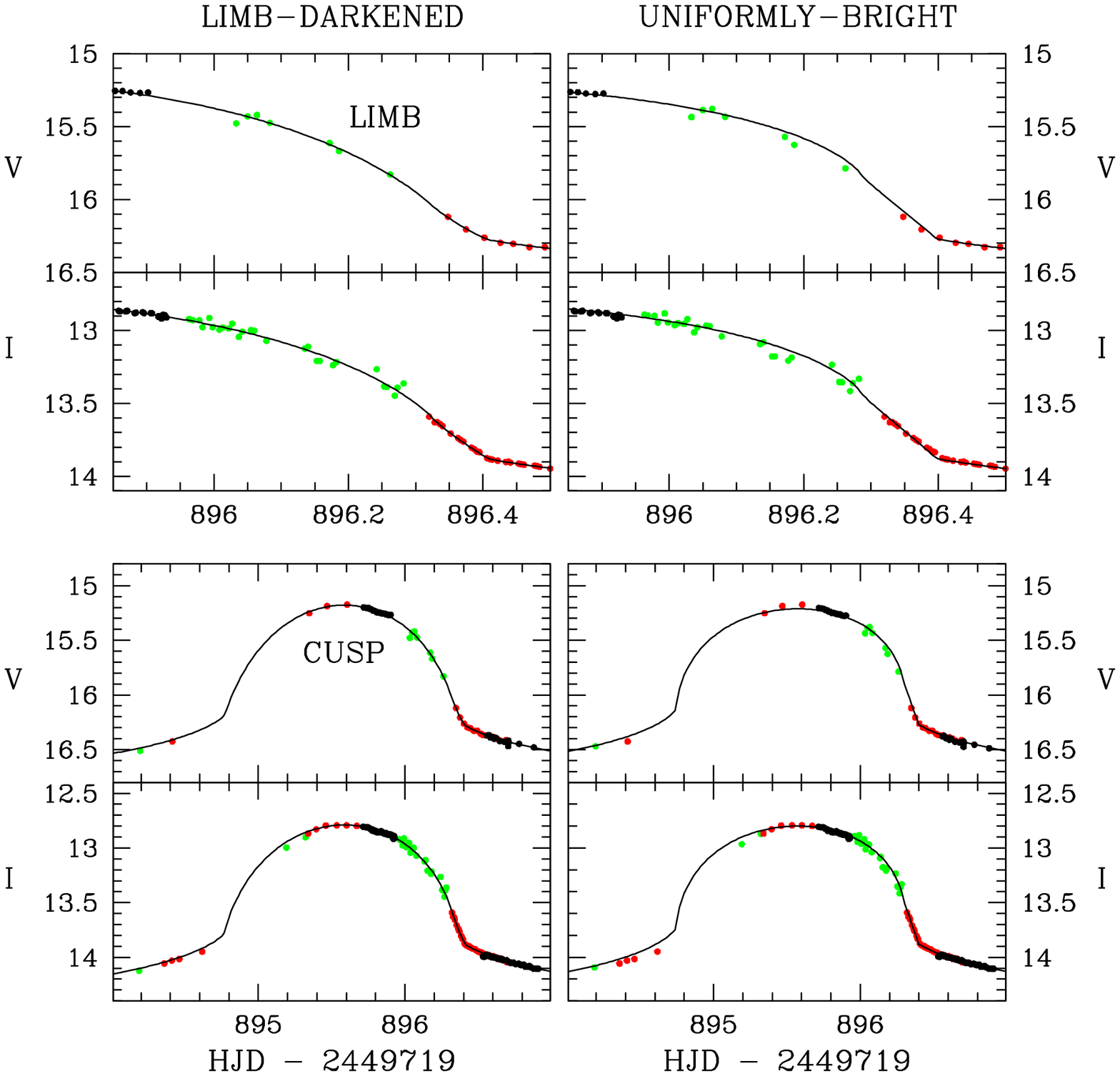}

\newpage

\vglue 1cm
\hglue -1cm
\epsfxsize=16cm\epsffile{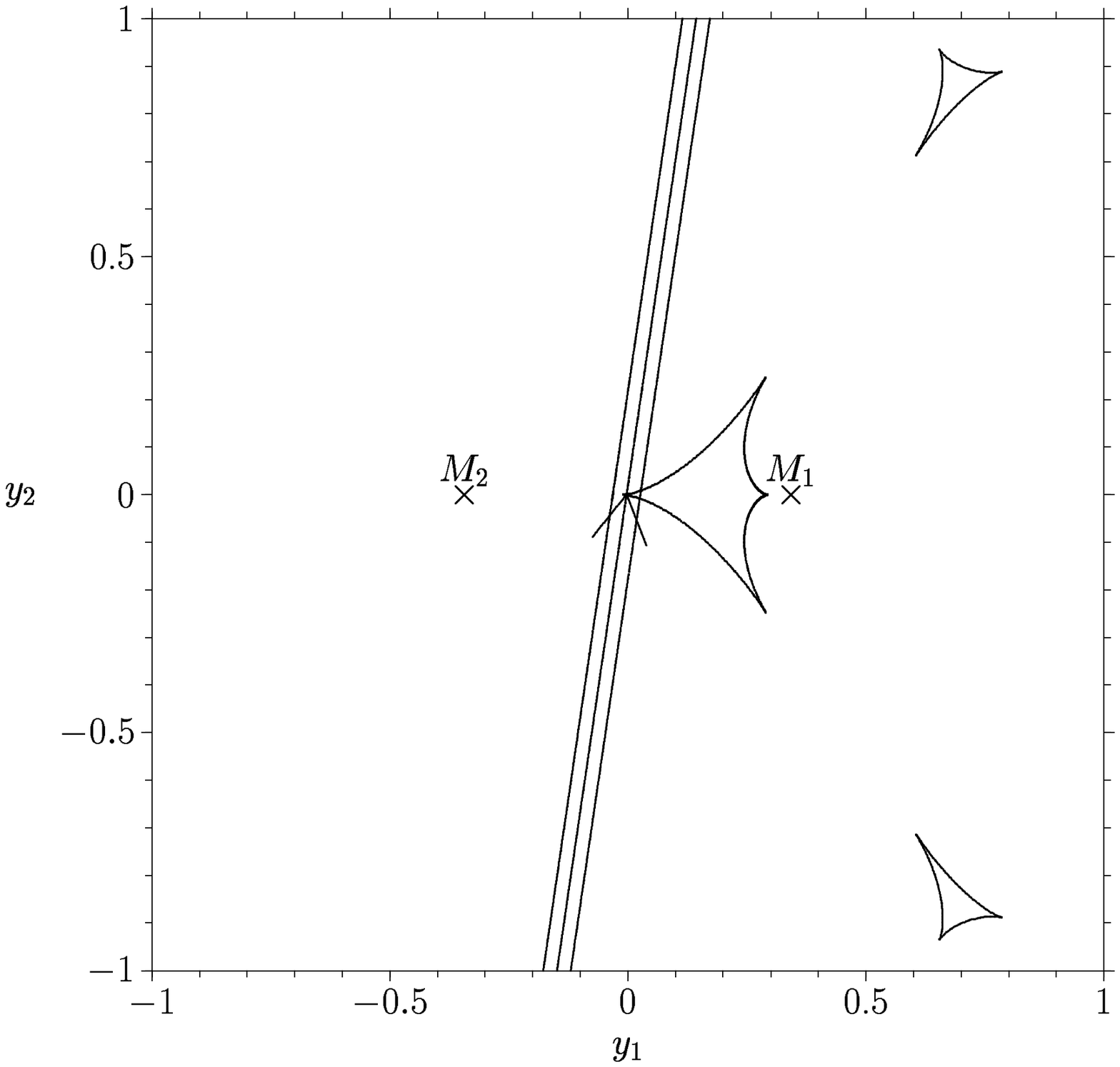}

\newpage

\vglue 1cm
\hglue 2cm
\epsfxsize=10cm\epsffile{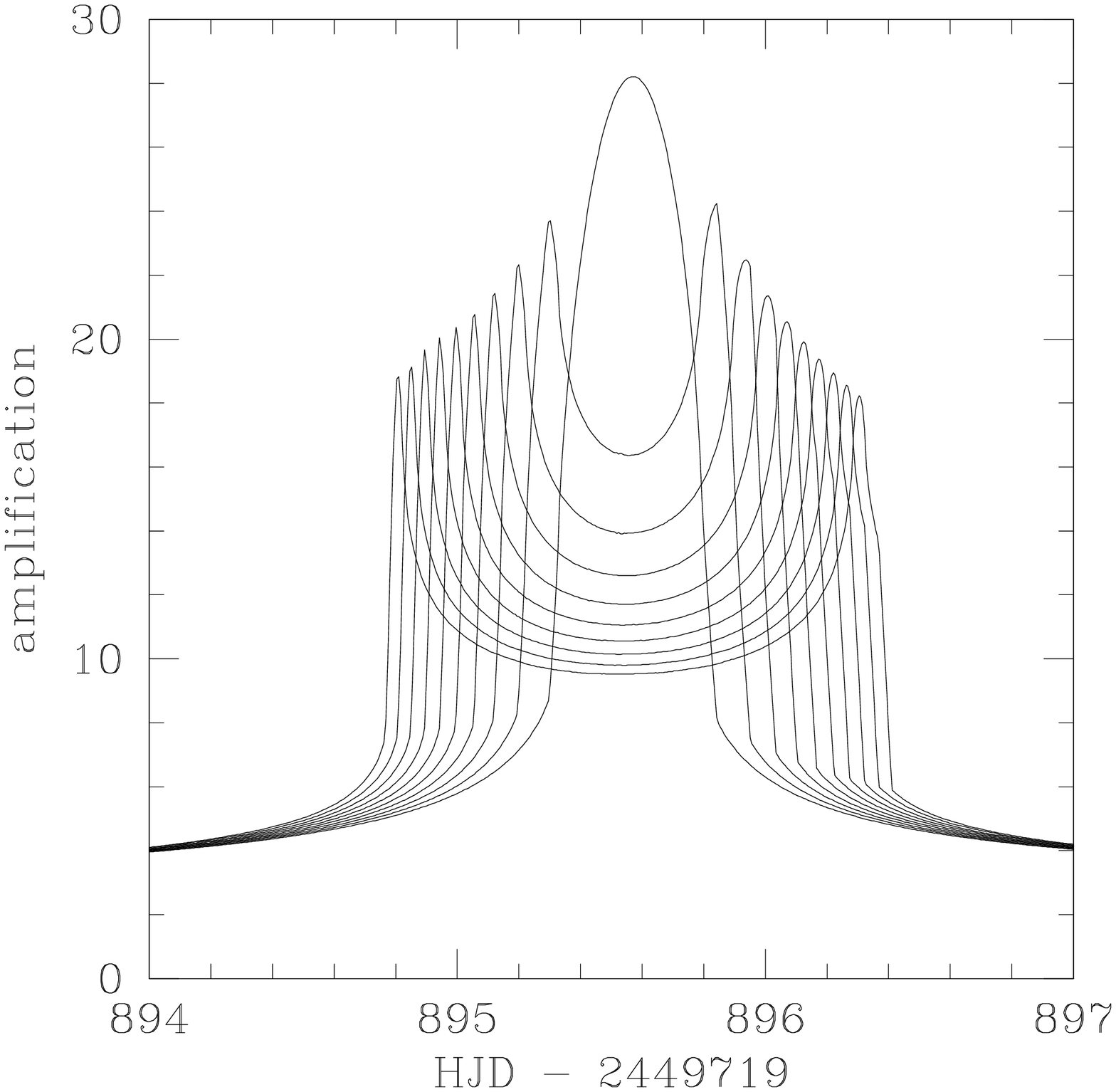}

\vglue -0.5cm
\hglue 2cm
\epsfxsize=10cm\epsffile{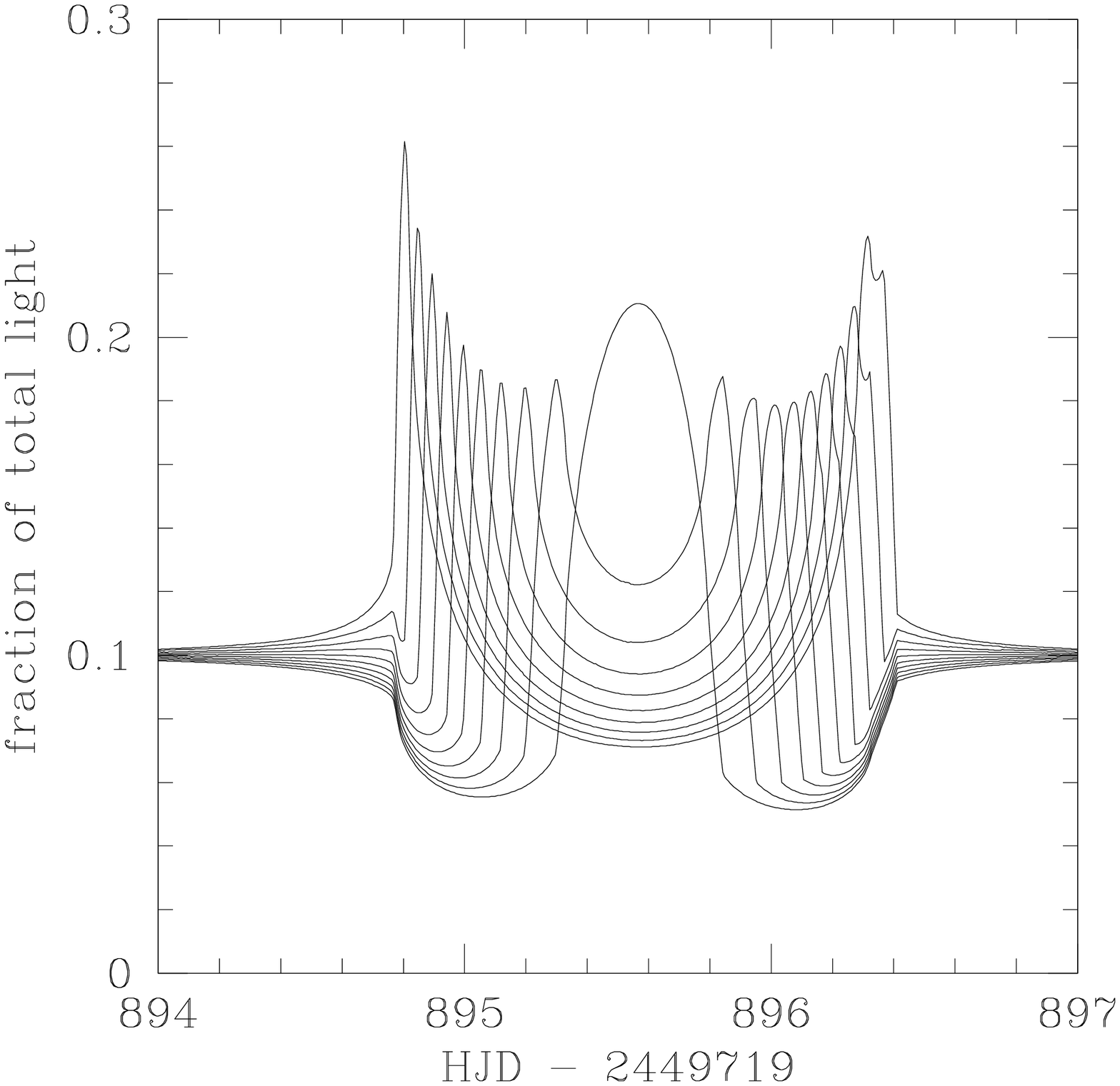}

\newpage

\vglue 1cm
\hglue -2cm
\epsfxsize=18cm\epsffile{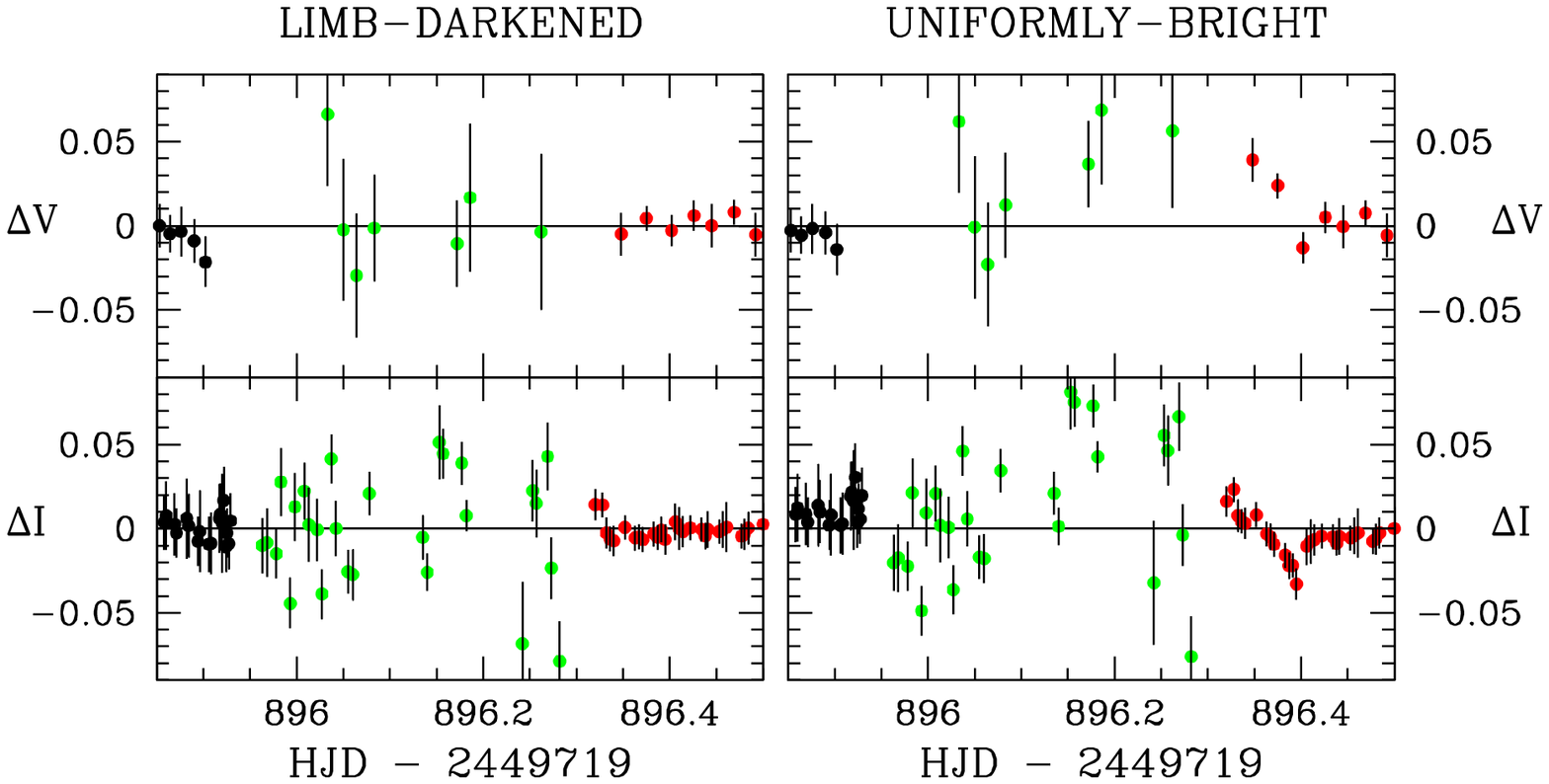}

\newpage

\vglue 1cm
\hglue -1cm
\epsfxsize=15cm\epsffile{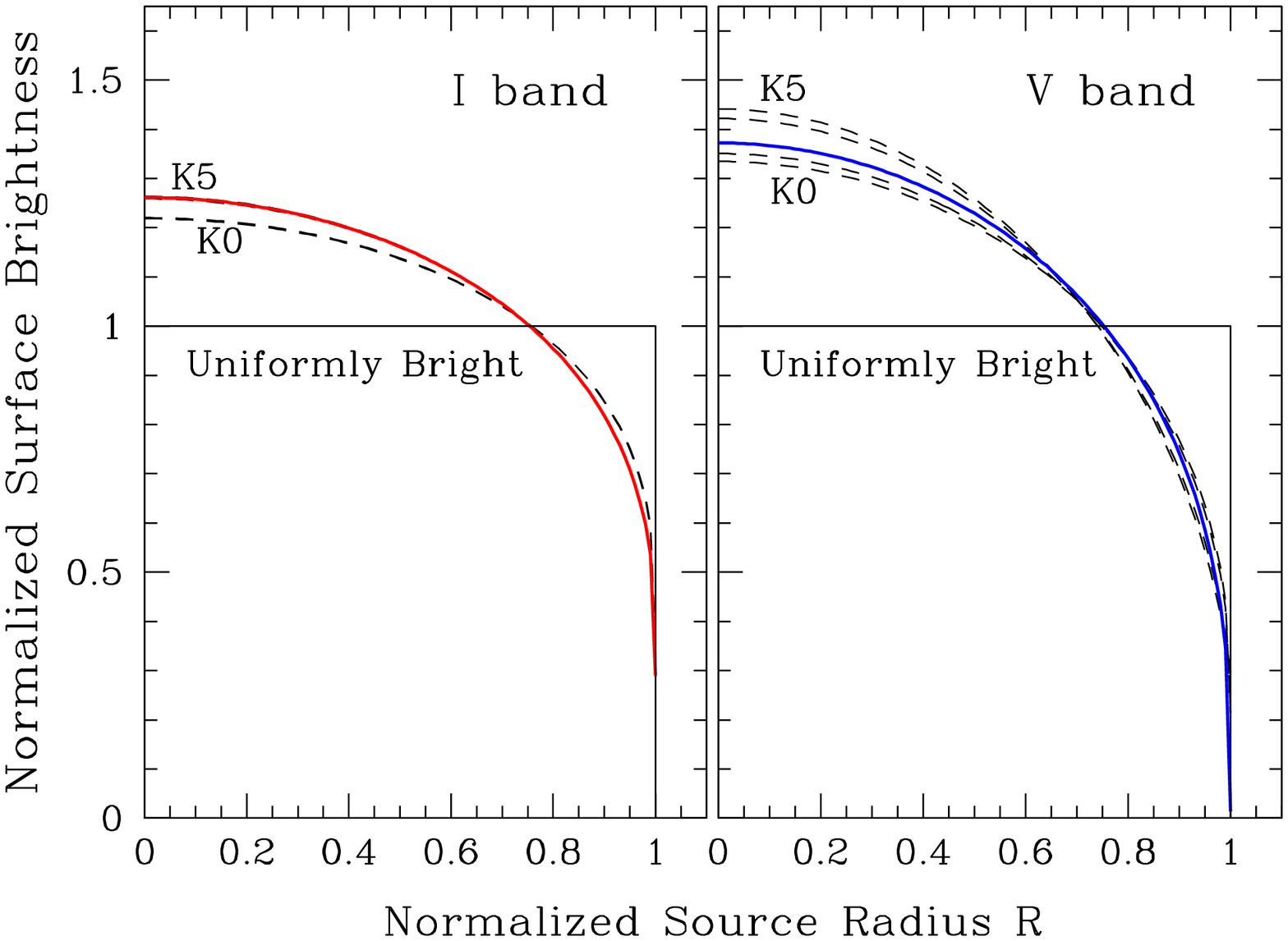}

\newpage

\vglue 1cm
\hglue -1cm
\epsfxsize=15cm\epsffile{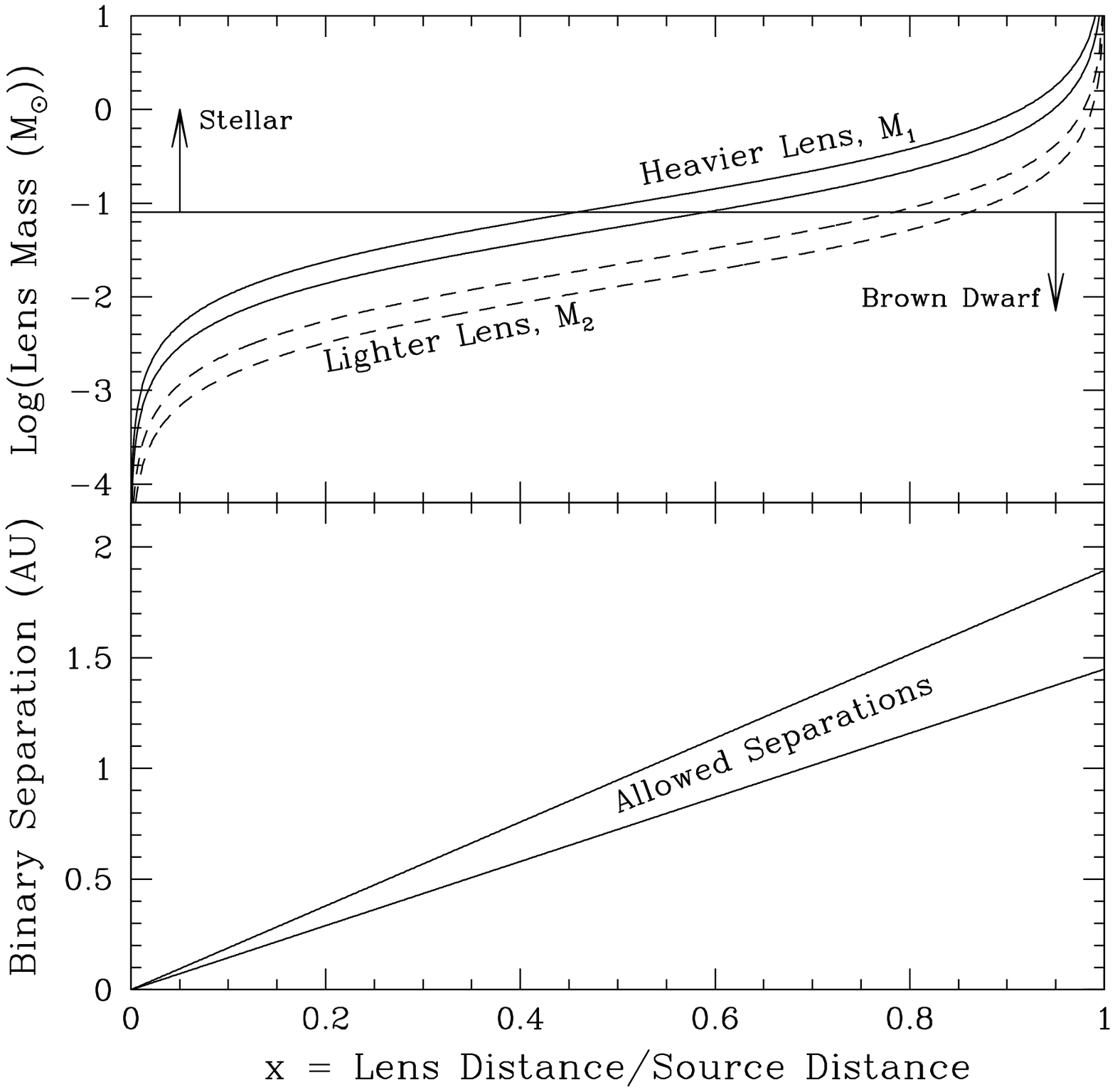}


\end{document}